# Reaction/Diffusion Competition Drives Anomalous Relaxation of Vitrimers

Makayla R. Branham-Ferrari[a], Shinian Cheng[b], Alexei P. Sokolov[b,c], and David S. Simmons[a]*

a. Department of Chemical, Biological, and Materials Engineering, University of South Florida, Tampa, FL, United States of America.
b. Department of Chemistry, University of Tennessee, Knoxville, TN, United States of America
c. Chemical Sciences Division, Oak Ridge National Laboratory, Oak Ridge, TN, United States of America.
* dssimmons@usf.edu

**Abstract:** Since their discovery in 2011, vitrimers – covalent associative network polymers – have challenged the traditional understanding of soft matter relaxation dynamics: unlike in typical glass-forming liquids, vitrimers' viscous relaxation can be entirely decoupled from their underlying structural (segmental) dynamics. Beyond this fundamental mystery, the origin of vitrimers' Arrhenius viscosity in the presence of super-Arrhenius structural relaxation behavior has been of high interest due to vitrimers' potential to provide readily reprocessable high-performance plastics. Here, we combine simulations, theory, and experiments to establish a foundational understanding of vitrimer relaxation dynamics. We identify two types of transient networks based on the ratio of atomic displacement scales required for bond exchange to those required to relax a segment. In systems where bond exchange only requires sub-segmental motion, we show that network relaxation is governed by a competition between chemical exchange reactions and segmental diffusion. This competition produces vitrimers' signature network/segment decoupling, while also driving a crossover between Arrhenius and super-Arrhenius behavior that is observed for many vitrimers. This work provides an explanation for longstanding puzzling features of vitrimer dynamics and establishes a foundation for rational vitrimer design.

Vitrimers are an emerging class of crosslinked polymer networks in which dynamic covalent bonds can exchange and rearrange, without any significant change in population of bonded states.[1] Because these bonds can rearrange at high temperatures without disintegration of the network, vitrimers could combine the mechanical stability of traditional permanently crosslinked networks with the recyclability of glasses and physical networks. Unlike classical transient/physical networks, the rate of vitrimer relaxation is decoupled from the mobility of their polymer segments. This challenges our fundamental understanding of how dynamic covalent networks relax and deform and has drawn proposed parallels to highly Arrhenius network glass-forming liquids.[2] Here we combine simulations, theory, and experiments to show that vitrimer dynamics are governed by a reaction-diffusion competition leading to emergence of two very different regimes: (i) a high temperature regime controlled by the local bond-exchange reaction rate and (ii) a low temperature regime controlled by segmental diffusion. This fundamental difference from traditional physical networks and glass-forming liquids explains why vitrimers can exhibit Arrhenius mechanical relaxation dynamics despite exhibiting super-Arrhenius segmental (glassy) relaxation dynamics, providing a foundation for their rational design.

Intuitively, polymer segments must relax for polymer network bonds to rearrange. Indeed, classical transient networks (e.g. ionic or hydrogen bonded networks), obey this expectation. In these systems, the network relaxation time $\tau_{bond}$ always exhibits a higher activation energy than does the $\alpha$ process via which segments relax. In such systems, the activation barriers $F_{ex}$ controlling local chemical exchange and $F_\alpha$ controlling $\alpha$ relaxation are additive:[3–5]

$$\tau_{bond} \sim \tau_\alpha \exp(F_{ex}/kT) = \tau_0 \exp\left(\left[F_{ex} + F_\alpha\right]/kT\right) \quad (1)$$

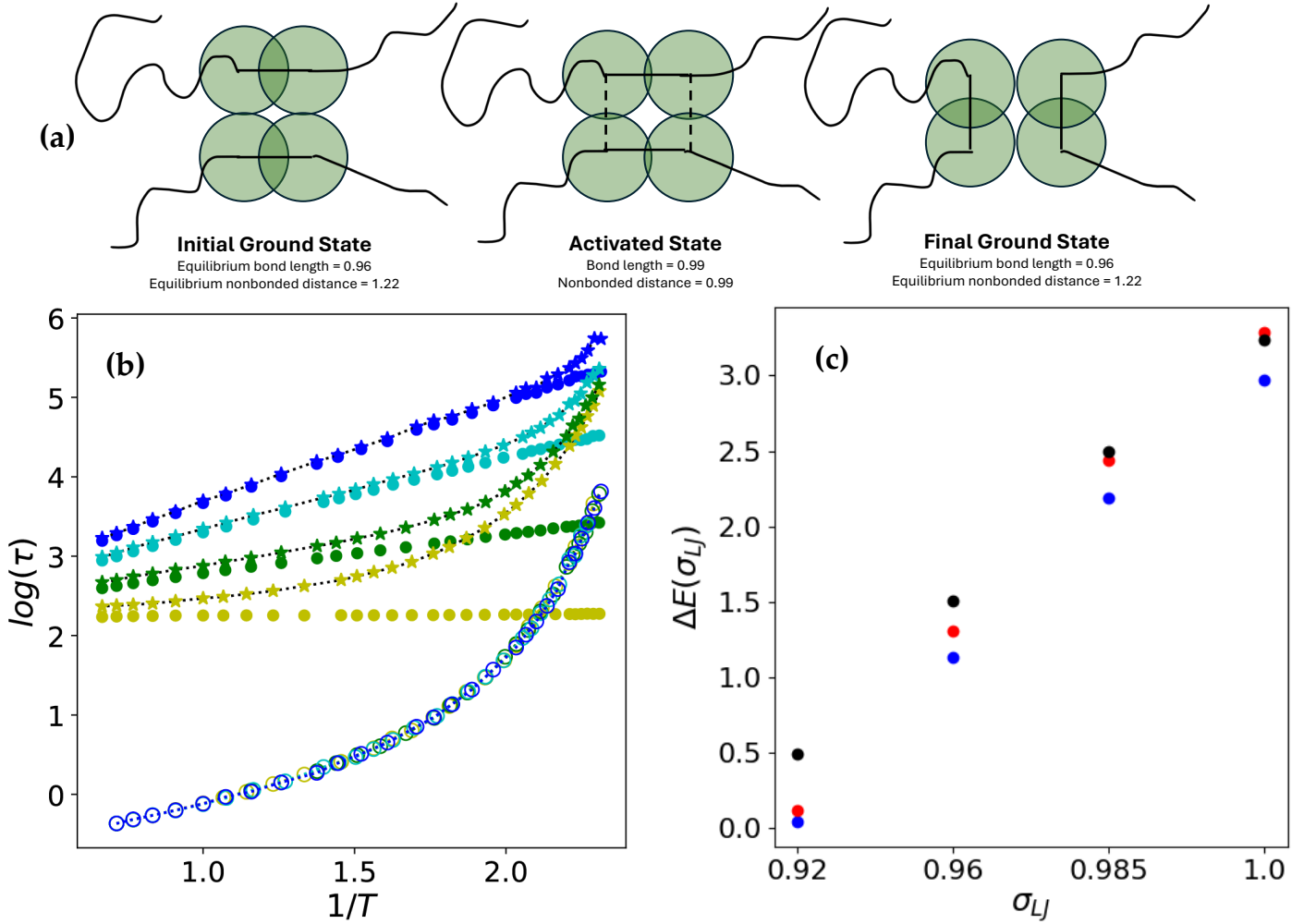

*Figure 1. (a) Schematic of geometry of bond exchange activated state for this model. We note that the activated state can also be accessed from a tetrahedral configuration of the four participating beads; this alters the activation entropy, but not the activation energy, deduced from the model. (b) Bond exchange timescales (filled circles), bond relaxation times (stars), and α-relaxation times (open circles) vs inverse temperature, for systems with exchangeable bead diameter $\sigma_{LJ}$ equal to 1 (blue), 0.985 (teal), 0.96 (green), and 0.92 (yellow). (c) Activation barriers for the bond exchange process vs $\sigma_{LJ}$ as predicted by theory (black points) and as obtained from Arrhenius fits to the high temperature regime (red) and full temperature range (blue) of $\tau_{ex}$ as reported in panel (b).*

Because the α process exhibits a super-Arrhenius temperature dependence over a broad temperature range, network relaxation is generally super-Arrhenius in these systems.[6–8]

Vitrimers commonly defy this intuition. Many exhibit an Arrhenius temperature dependence of viscosity and terminal relaxation time,[1,9] in some cases with an activation barrier lower than that of the α process.[10–12] Even more surprisingly, some vitrimers exhibit a transition on cooling from Arrhenius to super-Arrhenius network dynamics[11,13]. This suggests a fundamental change in the mechanism of bond rearrangements and its connection to α relaxation. These observations are inconsistent with the scenario encoded in eq. (1), wherein the α process sets the attempt time for bond rearrangement. The precise molecular mechanism controlling this behavior is a longstanding puzzle that hinders rational vitrimer design.

Several groups have argued that network relaxation in vitrimers can instead be described by the equation[4,14–16]

$$\tau_{bond} = \tau_{ex} + c\tau_\alpha, \quad (2)$$

where $\tau_{ex} = \tau_0 \exp(F_{ex}/kT)$ is a chemical timescale for bond exchange, and $c\tau_\alpha$ reflects some multiple of the α relaxation time $\tau_\alpha$.[3] However, the physical origin of equation (2) remains unsettled. It has variously been argued to emerge from a large entropic contribution to the activation barrier for bond rearrangement[4], or from a "2-step" scenario in which $\alpha$ relaxation must occur prior to bond dissociation or reassociation.[14] Conclusive molecular evidence for either proposal remains unavailable. It is also unknown why dynamic networks sometimes obey equation (1), and sometimes equation (2).

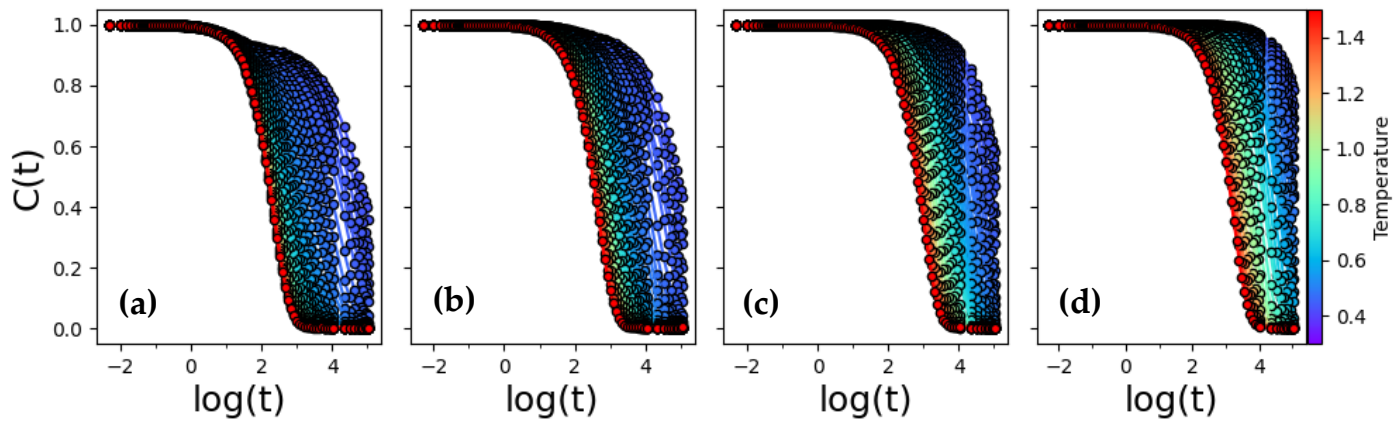

*Figure 2. Fraction of remaining original bonded partners (of re-associable bonds) at a time s+t after a start time s, averaged over multiple s, for systems with $\sigma_{LJ}$ equal to (a) 0.92, (b) 0.96, (c) 0.985, and (d) 1.*

We begin by performing hybrid Monte Carlo / molecular dynamics simulations of polymers with a highly local associative bond exchange mechanism (see methods for details). This system's bond exchange timescale $\tau_{ex}$ closely obeys an Arrhenius rate law, despite super-Arrhenius behavior of $\tau_\alpha$ (Figure 1b). Moreover, the activation energy for bond exchange is considerably less than the apparent activation energy of the $\alpha$ process. Thus this system reflects the puzzling behavior of many experimentally studied vitrimers by violating eq. (1).[10–12]

To understand the relationship between bond exchange and $\alpha$ relaxation in this simulated system, we develop a theoretical model for its bond exchange activation state (Figure 1a, and SI). This model reveals that the activation barrier for bond exchange in this simulation emerges from a mismatch between bonded and nonbonded length scales (Figure 1c). The model predicts an activation energy of 3.24 in reduced LJ units (Figure 1c), in excellent accord with the bond exchange activation energy for this simulated system. Guided by this model, we systematically reduce the non-bonded interaction length scale $\sigma_{LJ}$ between the beads participating in exchangeable bonds, bringing it closer to the bonded length scale, thereby tuning the activation barrier (Figure 1b)(see SI). The measured activation barrier appears in good agreement with predictions (Figure 1c). *These results clearly demonstrate that $\tau_{ex}$ is governed by the local exchange (i.e. chemical) activation barrier, without a significant contribution from the barrier to $\alpha$ relaxation.*

How can bond exchange occur without segmental relaxation? We postulate that this occurs when the displacement required for bond exchange is appreciably smaller than that of $\alpha$ relaxation. Based on the activated state geometry shown in Figure 1a (see SI), the largest displacement separating the ground state from the activated state for a bond exchange event is 0.13 $\sigma$, equal to the difference in distances between the non-bonded (1.12 $\sigma$) and activated (0.99 $\sigma$) states. By comparison, segments in this system can on average access displacements of 0.195 $\sigma$ through local rattling motions without engaging in an $\alpha$ process (as measured by the Debye-Waller factor – see SI). Because this is greater than the displacement required for bond exchange, particles can readily access the bond exchange activated state, via local rattling, without overcoming the $\alpha$ barrier. *This suggests that bond exchange can exhibit a lower activation barrier than the $\alpha$ process when bond exchange requires smaller displacements than $\alpha$ relaxation.*

The low temperature behavior of these systems (Figure 1b) raises another critical question: how can $\tau_{ex}$ be smaller than $\tau_\alpha$? In simulation, $\tau_{ex}$ counts *all* bond exchange events, including those that merely return the network to a prior state and therefore do not contribute to network relaxation. Experiments, however, probe bond or network decorrelation, to which recursive exchanges do not contribute. We therefore define a decorrelation function based on the fraction of beads in exchangeable bonds that retain, at a time $t$, their original bonded neighbor from some earlier arbitrarily chosen time 0. As shown by Figure 2, at low temperatures where $\tau_{ex} < \tau_{alpha}$, this function evolves a two-step character, wherein an initial rapid process of bond decorrelation is followed by a much slower, more stretched process. Could this be a hint to the origin of the surprising crossover noted above?

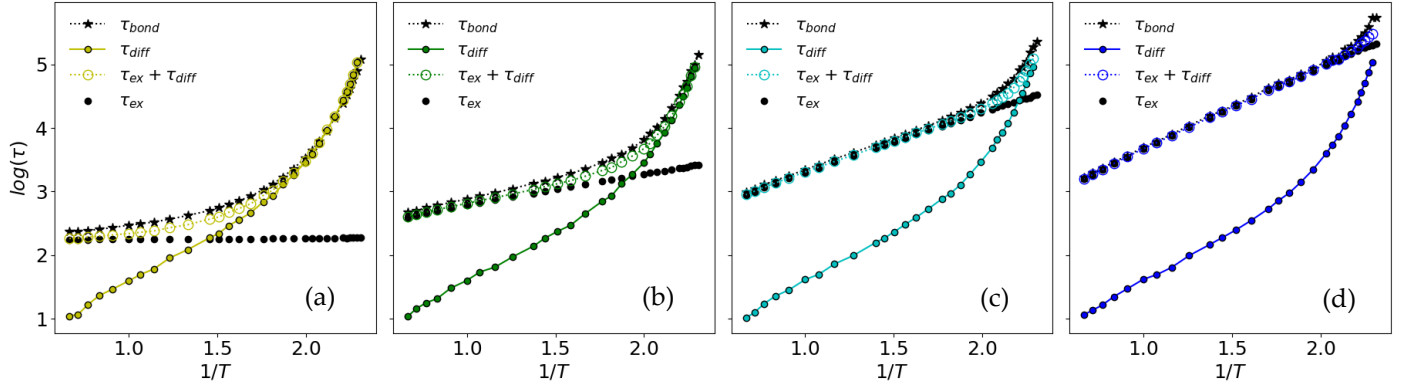

Figure 3. Chemical bond exchange time $\tau_{ex}$ (black circles), bond relaxation time $\tau_{bond}$ (black stars), timescale $\tau_{diff}$ at which segments' mean square displacements are $10^{1/2} \sigma$ (color-filled circles), and equation (4) prediction ($\tau_{ex} + \tau_{diff}$) of $\tau_{bond}$ (open circles) vs inverse temperature for systems with $\sigma_{LJ}$ equal to (a) 0.92, (b) 0.96, (c) 0.985, and (d) 1.

To answer this question, we develop in the SI a reaction/diffusion theory predicting this behavior. This theory captures two distinct events required for bonds to decorrelate: bonds must exchange (reaction) and move apart (diffusion). At high temperatures, diffusion is fast, and bond relaxation is limited by the reaction rate. Thus high-T bond relaxation is governed by chemical kinetics alone (unaffected by the $\alpha$ process) and is expected to obey a simple exponential, as we observe at high temperature in Figure 2.

At low temperature, where diffusion is much slower than reaction, bond relaxation is predicted to bifurcate into two time-regimes. The earlier regime involves exchange only of bonds that are initially proximate to one another. But these local exchanges prior to diffusion are entirely recursive. At longer times, diffusion occurs, allowing bond relaxation to resume. This predicts a low-temperature regime wherein bond relaxation obeys the following time dependence:

$$f(t) = \left[1 + A\left(e^{-\frac{t}{\tau_1}} - 1\right)\right] \exp\left[-\left(\frac{t}{\tau_{diff}}\right)^{\beta}\right]. \quad (3)$$

Here the relaxation time $\tau_{diff}$ is a subdiffusional timescale over which exchangeable groups translate together and apart. At low temperature, $\tau_{bond} \cong \tau_{diff}$ since $\tau_{diff}$ sets the terminal bond relaxation time.

Eq. (3) predicts the low-temperature emergence of a two-step bond relaxation process observed in Figure 2. Fitting the data in Figure 2a to eq. (3)(at low temperatures) or a simple stretched exponential (at high temperatures) allows extraction of $\tau_{bond}$. This timescale exhibits a crossover from Arrhenius behavior at high temperatures to super-Arrhenius at low temperatures (Figure 3), consistent with recent experiments and simulations.[11,13]

This two-regime behavior is predicted by the reaction/diffusion theory, which predicts $\tau_{bond} \sim \tau_{ex}$ at high temperature, and $\tau_{bond} \sim \tau_{diff}$ at low temperature. Employing the standard approach of approximating the crossover region by the sum of the asymptotics within this theory gives

$$\tau_{bond} \cong \tau_{ex} + \tau_{diff} \cong \tau_{ex} + c\tau_{\alpha} \quad (4)$$

for the reaction/diffusion case. Equation (4) is superficially comparable to equation (2), but the origin of $c\tau_{\alpha}$ as an approximation for $\tau_{diff}$ signifies that reaction/diffusion competition, rather than direct coupling to the $\alpha$ process, governs the observed physics.

In addition to predicting the shape of the bond decorrelation function, this reaction/diffusion theory provides via eq. (4) an excellent description of temperature-dependent dynamics in our simulated systems (Figure 3). It correctly captures the high and low temperature limiting behaviors of the systems: at low T,

$\tau_{bond}$ converges to the (subdiffusive) timescale on which a typical segment displaces the average distance between exchangeable bonds (around 2 σ); at high T, $\tau_{bond}$ converges to $\tau_{ex}$. Overall, equation (4) is found to predict $\tau_{bond}$ to leading order. Modest deviations observed in the crossover regime are expected given that the crossover was approximated in eq. (4) from a sum of asymptotics.

This reaction-diffusion scenario is expected only when the activated state for bond exchange can be accessed via sub-α rattling motions. This suggests that transient networks fall into two types (or at least a spectrum with two limits). A system's type is controlled by the value of its displacement ratio $x = r^*_{ex}/r^*_\alpha$, where $r^*_{ex}$ is the displacement required to access the bond exchange activated state and $r^*_\alpha$ is the displacement required for a segmental α relaxation event. When $x \geq 1$, chemical bond exchange locally requires an α event and equation (1) approximately holds (type I). When, $x < 1$, reactions involve very small atomic displacements and the reaction-diffusion scenario (equation (4)) holds (type II).

Type II systems are characterized by unique features. Here, the activation barrier of network relaxation can be less than that of the α process. As shown in prior work, the relaxation time for the network can still be appreciably larger than $\tau_\alpha$ when bond exchange involves a large activation entropy (large steric factor[16]). As per eq. (4), this leads to scenarios in which $\tau_{ex}$ is Arrhenius and insensitive to $\tau_\alpha$ at high T; these systems become super-Arrhenius only when $\tau_{diff}$ ($c\tau_\alpha$) and $\tau_{ex}$ become comparable. In contrast, type I systems exhibit none of these features: their $\tau_{bond}$ obeys eq. (1), exhibits a stronger temperature dependence than $\tau_\alpha$, and therefore exhibits a super-Arrhenius temperature dependence in the glass formation range.

To probe this behavior experimentally, we measure segmental and terminal relaxation times in a series of vitrimers with imine bonds based on amine-terminated telechelic poly(propylene glycol) (PPG) crosslinked by benzene-1,3,5-tricarbaldehyde (structure shown in Figure 4a), with varying PPG molecular weight from 4000 g/mol to 230 g/mol (see SI for details). Their terminal relaxation time is expected to track with bond decorrelation times as probed in simulations.

As seen in Figure 4a, network relaxation in these systems is insensitive to $\tau_\alpha$ at high T: $\tau_\alpha$ varies dramatically with PPG length between crosslinks, while network relaxation varies little. At lower temperatures, we note the beginning of a crossover to super-Arrhenius network relaxation, with timescale sensitive to $\tau_\alpha$ (Figure S6). These behaviors are consistent with features of type II (reaction-diffusion) systems described above. Moreover, the temperature—dependent network relaxation time of these systems is well described by equation (4) (Figure S6).

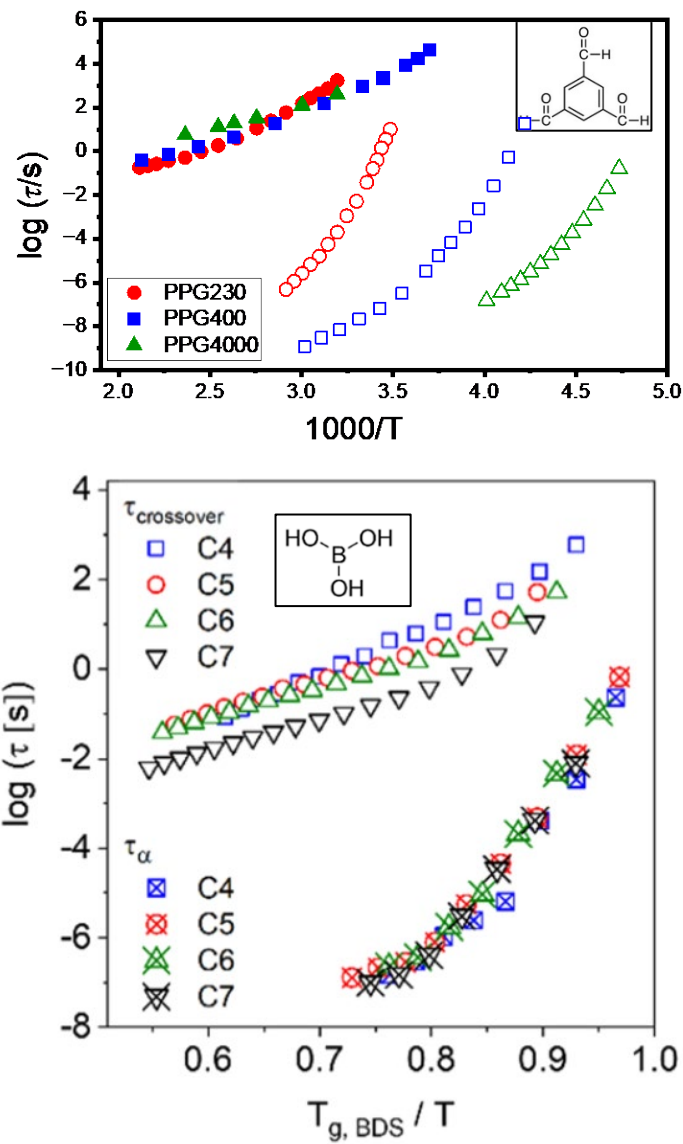

*Figure 4. (top) The same quantities plotted vs inverse temperature for imine-crosslinked vitrimers experimentally probed in this study. Inset shows the crosslinker chemistry in each case. (bottom) Network relaxation time ($\tau_{crossover}$) and α relaxation times vs scaled inverse temperature for a series of boronic ester crosslinked ethylene vitrimers, reproduced with permission from Soman et al.[10]*

A similar analysis can be applied to the recently studied vitrimers of Evans and coworkers.[11] Their systems exhibit a pronounced super-Arrhenius upturn of $\tau_\alpha$ around $T_g/T \sim 0.8$, yet the network relaxation process remains Arrhenius to appreciably lower temperatures (Figure 4b). The super-Arrhenius upturn of the network relaxation process instead occurs at lower temperatures, where $\tau_\alpha$ begins approaching the network relaxation time – again indicating the expected behavior of type II systems.

Puzzlingly, based on where the network relaxation process becomes super-Arrhenius, the constant c in the equation 4 becomes quite large for these experimental systems – of loose order $10^7$ - $10^{10}$ (Figure S6). The aforementioned simulated system is reasonably consistent with Rouse theory for an unentangled network, in which we expect $c \sim n^2$ (where $n$ is the number of segments per primary chain). The values of $c$ in the experimental systems far exceed this value. The origin of this large value is an open and important question, as it governs the crossover from Arrhenius to super-Arrhenius network dynamics. There is precedent for large kinetic prefactors in bond lifetime renormalization in sticky entangled networks or in large chemical entropic prefactors, but neither scenario seems consistent with this case.

These findings establish a fundamental understanding of why vitrimer dynamics can be so uniquely decoupled from their $\alpha$ (glassy) relaxation process. The small local displacements required for bond exchange in many covalent exchange reactions permit $\alpha$ relaxation and bond exchange to occur independently – a scenario radically different from nearly all other network systems, where the $\alpha$ process is the fundamental precursor to all other structural rearrangement processes.[2] Thus even when vitrimer bond exchange is much slower than $\alpha$ relaxation, bond exchange does not causally depend on the $\alpha$ and therefore does not inherit the $\alpha$'s temperature dependence. Instead, the two processes obey a reaction/diffusion scenario wherein the slower process serves as the singular gating process for relaxation and uniquely determines its temperature dependence.

This understanding points to a new approach for rational control of transient polymer networks, by selecting crosslink chemistry to modulate the spatial displacement required to access the activated state for bond exchange. Similarly, the activation entropy of the chemical exchange is central to yielding network relaxation that is much slower than the segmental relaxation time at high temperatures, despite the lower activation enthalpy of bond exchange.[4,16,17] The Arrhenius/super-Arrhenius crossover, and thus the combined thermal stability and reprocessability of the material, can then be tuned by varying these factors. The proposed scenario provides a clear qualitative picture of parameters controlling viscoelastic properties and network rearrangements in vitrimers and resolves the longstanding apparent contradiction between predictions of equations 1 and 2.

## Data Availability

Simulation and experimental data will be made publicly available prior to publication via the University of South Florida Digital Commons Site.

## Acknowledgements

MRB and DSS acknowledge funding support from the Air Force Office of Scientific Research under grant number FA9550-22-1-0427. SC and APS acknowledge financial support from NSF Polymer program (award DMR-2515834).

## Author Contributions

MRB performed and analyzed all simulations under DSS' supervision. SC performed and analyzed all experiments under APS' supervision. MRB drafted the manuscript; all authors contributed to editing and revision. DSS and APS conceived the research.

## Competing interests

The authors declare no competing interests.

## Additional Information

Extended data, extended methods are available in supporting information file.

# Methods

*Simulations*

Simulated systems are comprised of end-linked primary strands, connected by linear (functionality 2) linkages. The general simulation model is adapted from a standard bead-spring model, wherein non-bonded beads interact via an attractive 12-6 Lennard-Jones potential and bonded beads additionally interact via a Finitely Extensible Nonlinear Elastic bonding potential.[18] Simulations are performed within the Large-scale Atomic/Molecular Massively Parallel Simulator (LAMMPS).[19] Each simulated system consists of 3,000 primary bead-spring polymer strands, each containing 8 non-reactive beads, and with 1 bead on each end participating in an exchangeable bond.

Nearby bead that are members of exchangeable bonds can exchange partners through a Metropolis Monte Carlo bond exchange move widely employed to accelerate conformational annealing of long chains[20], and recently employed for vitrimers[21]. The exchangeable groups in this model have no preferential affinity for each other and thus have no tendency to aggregate. Data on system dynamics are collected over a range of temperatures cooling towards the glass transition via an established annealing procedure[22]. Additional methodological details can be found in Electronic Supplementary Information.

*Simulation Analysis*

Simulation analysis employs the publicly available Amorphous Molecular Dynamics Analysis Toolkit package developed by the Simmons group.[23] Segmental $\alpha$ relaxation dynamics of the polymer are computed from the self-part of the intermediate scattering function computed at a wavenumber near the first peak in the structure factor. We then obtain a relaxation time from by fitting the long-time portion of this response to a stretched exponential function, with $\tau_\alpha$ defined as the timescale at which this function interpolates to a value of 0.2.[22] We compute the mean-square displacement of the system in the standard manner. Mean square displacements vs time for the lowest temperature of each simulated system are shown in Figure S 2. We then define the Debye-Waller factor as the mean-square displacement at a time of about 1 $\tau_{LJ}$.[22,24]

*Theoretical model of bond-exchange activated state in simulation model*

The activated state in this model can be understood by reference to Figure 1a in the main text. The energy of the activated state can be reasonably approximated as a sum of four energies: the two bonded and two nonbonded adjacent pairs shown in the activated state of Figure 1a. Denoting the distances between each adjacent pair of bonded beads as $l_b$ and between each adjacent pair of nonbonded beads as $l_{nb}$, the energy of this state is given by

$$E^* = 2E_{LJ}(r = l_{nb}) + 2E_{FENE}(r = l_b) \quad (5)$$

By symmetry, the lowest-energy activated state occurs when $l = l_{nb} = l_b$, such that its energy is given by

$$E^* = 2E_{LJ}(l) + 2E_{FENE}(l) \quad (6)$$

The activated state is then found by minimization of equation 4 with respect to l, i.e. $dE^*/dl = 0$, yielding.

$$0 = -\frac{4\varepsilon_{LJ}}{l^*}\left[12\left(\frac{\sigma_{LJ}}{l^*}\right)^{12} - 6\left(\frac{\sigma_{LJ}}{l^*}\right)^{6}\right] + \frac{Kl^*}{1-(l^*/R_0)^2}$$
$$-\frac{4\varepsilon_{FENE}}{l^*}\left[12\left(\frac{\sigma_{FENE}}{l^*}\right)^{12} - 6\left(\frac{\sigma_{FENE}}{l^*}\right)^{6}\right] + \varepsilon \quad (7)$$

Numerical solution of equation (7) yields $l^* = 0.995$ and $E^* = 3.24$ for the standard model parameters. As shown in figure 1c in the main text, with reduced $\sigma_{LJ}$ the model predicts a reduced activation barrier.

*Experiments*

Poly(propylene glycol) bis(2-aminopropyl ether) (PPG230, PPG400, and PPG4000 with Mn=230, 400, and 4000 g/mol, respectively) was purchased from Sigma-Aldrich, and benzene-1,3,5-tricarbaldehyde (BTA, 98%) was purchased from Tokyo Chemical Industry (TCI). All samples were used as received. Reaction of these samples formed crosslinked networks (Figure S3), as evident from the extended rubbery plateau (Figure S4), and have sufficient number of free $NH_2$ groups (0.25*2 per molecule) critical for transamination bond exchange mechanism.

Small Amplitude Oscillatory Shear (SAOS) rheological measurements in the angular frequency range of $10^{-3}$-$10^{2}$ rad/s were conducted in linear regime with the strain-controlled mode of an AR2000ex rheometer (TA instrument) utilizing parallel plate geometry with 8mm in diameter. Dielectric measurements in the frequency range from $10^{-2}$ up to $10^6$ Hz were performed using a Novocontrol GMBH Alfa impedance analyzer. In the high-frequency measurements ($10^6$ to $10^9$ Hz), we applied

Agilent 4291B impedance analyzer connected with the Novocontrol GMBH system. Additional experimental methodological data are found in the SI.

# Supplementary Information for "Reaction/Diffusion Competition Drives Anomalous Relaxation of Vitrimers"

Makayla R. Branham-Ferrari[a], Shinian Cheng[b], Alexei P. Sokolov[b], and David S. Simmons[a]*

**Extended Simulation Methods**

*Simulation forcefield and protocol*

Within the simulation model, non-bonded interactions are modeled via the 12-6 Lennard Jones potential:

$$E_{LJ}(r) = 4\varepsilon_{LJ}\left[\left(\frac{\sigma_{LJ}}{r}\right)^{12} - \left(\frac{\sigma_{LJ}}{r}\right)^{6}\right], \quad (1)$$

In general, all beads employ the standard parameters $\sigma_{LJ} = 1$ and $\varepsilon = 1$, except that the $\sigma_{LJ}$ parameter for non-bonded interactions between beads participating in exchangeable bonds is systematically reduced to values of 0.985, 0.96, and 0.92 as described in the main text. Only interactions between these beads are changed; interactions between and with other beads in the chain remain unaltered.

Bonded interactions are modeled via the Finitely Extensible Nonlinear Elastic (FENE) potential:

$$E_{FENE}(r) = -0.5 K R_0^2 \ln\left[1 - \left(\frac{r}{R_0}\right)^2\right] + 4\varepsilon_{FENE}\left[\left(\frac{\sigma_{FENE}}{r}\right)^{12} - \left(\frac{\sigma_{FENE}}{r}\right)^{6}\right] + \varepsilon, \quad (2)$$

with $K = 30$, $R_0 = 1.50$, $\varepsilon = 1.0$, $\sigma = 1.0$ for all bonds, including both exchangeable and non-exchangeable bonds.

Simulations employ a Verlet time integration algorithm with timestep 0.005 $\tau_{LJ}$ (where $\tau_{LJ}$ is the Lennard Jones unit of time an corresponds to approximately 1 ps in real units). Temperature and pressure are controlled with the Nose Hoover thermostat/barostat pair, as implemented in LAMMPS, with damping parameters 2 $\tau_{LJ}$ for both. All simulations are performed at constant pressure P = 0.

In each primary chain, the end beads participate in a exchangeable bond, which can be with any other end bead of any chain. The system is initialized with each chain end bonded with the other end of the same chain, such that the system initially consists of 3000 rings. Bond exchange then proceed through a long annealing process (see below) to achieve an equilibrium bond distribution.

Bond exchange events are performed via the bond/swap algorithm in LAMMPs, which behaves as follows. Using the *group* keyword, exchanges are restricted to the telechelic end beads in each chain. Each CPU core employed in executing simulation attempts a bond exchange every *Nevery* timesteps. In order to ensure that the rate of bond exchange events is held constant across simulations even if the number of processor cores *Ncores* is varied, we scale *Nevery* such that *Nevery*/*Ncores* = 1 timesteps/attempt. In a system with 3000 exchangeable bonds (1 per primary chain), this leads to a temperature-invariant attempt rate of (on average) 15 tau/(attempt/bond) (see Figure S 1. Exchange attempt timescale: time/(attempt/bond).). Exchange attempts are made only on bonds containing beads within a distance 1.5 σ of each other. Exchanges are accepted or rejected according to the Metropolis criterion.

Other simulations of vitrimers have commonly employed either much more complex MC moves or multibody forcefields[12] to simulate bond exchange. The MC move we employ here has the virtue of physical simplicity, is quite efficient (allowing access to long timescales), has a well-defined activated state (main text Figure 1a), and has a substantial activation entropy, making it an excellent minimal model for real covalent exchange reactions.

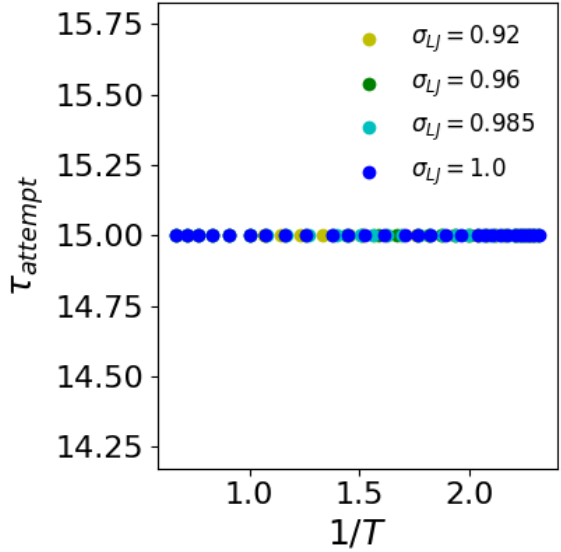

*Figure S1. Exchange attempt timescale: time/(attempt/bond).*

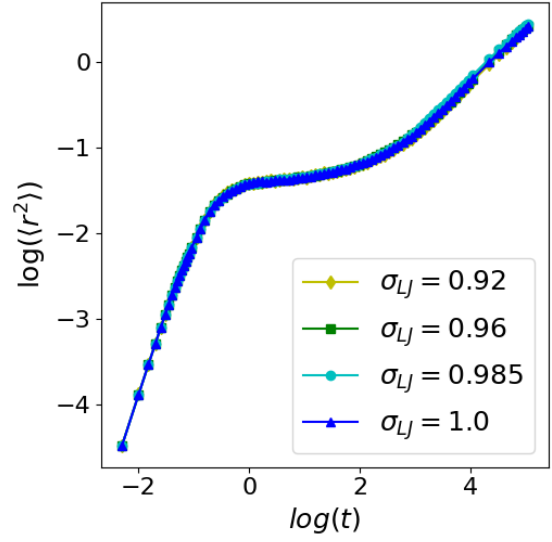

*Figure S2. Logarithm of mean square displacement plotted vs time for the lowest temperature of each simulated system, as shown in the caption.*

Model systems with reduced $\sigma_{LJ}$ for bonding beads are produced by modification of the standard system. An initial equilibrium configuration of the standard system is produced at T=1.5 via an isothermal equilibration period of $10^5$ $\tau_{LJ}$. During this period, reactive bonds exchange, allowing the topology of the system to evolve to equilibrium. Other systems are then produced by reducing $\sigma_{LJ}$ for bonded beads to the target value (shown in main text Figure 1c) and then subjecting the system to an addition $10^3$ $\tau_{LJ}$ isothermal equilibration period.

Equilibrium configurations for each system are then produced over a range of temperature while cooling towards $T_g$, employing the Predictive Stepwise Quench (PRESQ) algorithm that was described and validated in prior work[1]. In summary, this algorithm progressively quenches the system to lower temperatures, employing an isothermal isobaric annealing step at each temperature of sufficient length to exceed $10\tau_\alpha$. Because the structure of the bonded network itself is athermal, this is sufficient to yield equilibrium dynamics. Dynamical data is then collected from long runs at each temperature beginning with these equilibrated configurations.

*Simulation Analysis*

Segmental $\alpha$ relaxation dynamics of the polymer are computed from the self-part of the intermediate scattering function, defined as

$$F_s(q, \Delta t) = \frac{1}{HJN} \sum_{h=1}^{H} \sum_{j=1}^{J} \sum_{n=1}^{N} \exp\left[-i\boldsymbol{q}_h \cdot \left(\boldsymbol{r}_n(s_j + \Delta t) - \boldsymbol{r}_n(s_j)\right)\right] \quad (3)$$

where $r_n(t)$ is the position of particle $n$ at time $t$. $s_j$ is an arbitrary initial start time, $\boldsymbol{q}_h$ is a wavevector, N is the number of particles in the system, and J is the number of start times employed. To arrive at a mean $F_s$ for wavenumber $q$, we average over wavevectors $\boldsymbol{q}_h$ corresponding approximately to this wavenumber; H is the number of such wavevectors for a given wavenumber. We compute dynamics at a wavenumber of 7.07, corresponding approximately to the first peak in the structure factor. We obtain a relaxation time from the resulting function via a standard simulation approach: we fit the long-time portion of this response to a stretched exponential function, and we then define $\tau_\alpha$ is the timescale at which this function interpolates to a value of 0.2.

We compute the mean-square displacement of the system in the standard manner as follows.

$$\left\langle r(\Delta t)^2 \right\rangle = \frac{1}{SN} \sum_{j=1}^{J} \sum_{n=1}^{N} \left(\boldsymbol{r}_n(s_j + \Delta t) - \boldsymbol{r}_n(s_j)\right)^2 \quad (4)$$

Mean square displacements vs time for the lowest temperature of each simulated system are shown in Figure S 2. We then define the Debye-Waller factor as the mean-square displacement at a time of about 1 $\tau_{LJ}$, consistent with many prior simulation works.

**Reaction-diffusion model for bond exchange**

We propose that bond relaxation can be modeled as a reaction diffusion problem governed by the following set of pseudo-chemical equations:

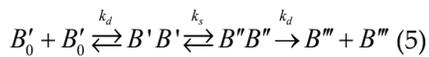 (5)

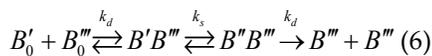 (6)

Here $B$ denotes an exchangeable bond. A subscript of '0' denotes that the bond is in a condition in which it is not geometrically proximate to another exchangeable bond, such that an $\alpha$-gated diffusive process would be required to bring it into a configuration in which an exchange can occur via a local chemical process. $B'$ denotes a bond that exists at an arbitrary start time $s$. $BB$ states denotes paired couplets of exchangeable bonds that are positioned proximate to one another such that they can engage in a local bond exchange event without requiring $\alpha$ relaxation. $B''$ denotes a bond that has been newly formed (after time $s$) by an exchange involving a $B'$ bond, and that has not diffused away from its original neighbor, such that it is possible for it to undergo a recursive exchange back to its original $B'$ state. $B'''$ denotes a bond that undergone an exchange after time $s$, and has additionally diffused away from its original exchange partner such that it is entropically improbable for it to exchange back to its original (at $t = s$) configuration.

The physics of these equations can be understood as follows. In equation (8), the first equilibrium process denotes pseudochemical representation of a diffusive process in which two $B_0'$ groups, initially far from one another, come together into a proximate configuration $B'B'$ via a diffusive process. This step is reversible, because the paired couplet can readily diffuse apart once again. $B'B'$ can also undergo a chemical reassociation process, to form a new pair of (permuted) bonds. Finally, these permuted bonds (denoted as $B''B''$) can either exchange back (to $B'B'$), or they can diffuse apart. The latter diffusional process is effectively irreversible, because reversal of this process (to allow the $B''B''$ state to be recovered) would require the *same* two $B$ groups that initially exchanged to come back together after diffusing apart – an entropically improbable event. We model two distinct limits of this scenario.

<u>At high temperature,</u> the diffusional process is much faster than the local chemical bond exchange process, such that the overall reaction-diffusion problem is in the reaction-limited regime. In this limit, the diffusive pseudo-reaction $B_0' + B_0' \overset{k_d}{\rightleftarrows} B'B'$ is always in local equilibrium. Moreover, all $B''B''$ and $B''B'''$ is effectively instantly destroyed via the diffusional process that creates $B'''$ groups. It follows that in this limiting case $[B] = [B'] + [B''']$ (where square brackets denote concentration), and that all complexes containing $B''$ are of zero concentration. One can then write the following rate equation for B' (where B' denotes any bond that existed at start time s, regardless of whether it is paired with another swappable bond):

$$\frac{d[B']}{d\Delta t} = k_s \left( -2[B'B'] - [B'B'''] \right) \quad (7)$$

Moreover, when diffusion is fast compared to reaction, at any given time the concentration of a diffusively reversible complex is given simply by the thermodynamic probability of this complex, such that

$$[B'B'] = a[B']^2$$
$$[B'B'''] = a[B'][B'''] \quad (8)$$

Within a lattice statistical mechanical model of the probability of adjacency, $c$ would be expected to be approximately the nearest neighbor lattice site count, if concentration is given in volume fraction. In this case, this would reflect the count of nearest neighbor exchangeable bonds that are in the correct position and orientation to allow for a bond swap without undergoing an $\alpha$ process. We initially take a ~ 1 given the steric difficulty of arriving in such a state (i.e. for each exchangeable bond there is only a single position, at a coarse lattice level, in which a second swappable bond can lie at any given time in which it is immediately swappable). This is expected to be chemically reasonable for exchangeable bonds that general have restrictive steric constraints for the location of a bond with which they can swap.

Solution of the above set of equations leads to an approximately exponential decay of bond correlations at high temperature, in the reaction-limited regime. This is consistent with the behavior observed in simulation.

*At low temperature*, the diffusional timescale becomes much longer than the local exchange reaction timescale due to the super-Arrhenius nature of the diffusional process in glass-forming liquids. In this regime, ultimate bond relaxation is diffusionally limited. Here, we model a separation of timescales between short-time reactions, which occur only between bonds that are initially adjacent to one another (i.e. that do not require any diffusional process to react), and longer-timescale diffusional processes. In the short time limit, a single reaction dominates the system, since diffusion cannot yet occur:

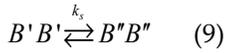

$$B'B' \underset{}{\overset{k_s}{\rightleftharpoons}} B''B'' \quad (9)$$

The kinetic equations for the change in [B'] can then be written as follows:

$$\frac{d[B'B']}{d\Delta t} = k_s [B''B''] - k_s [B'B'], (10)$$

and

$$\frac{d[B''B'']}{d\Delta t} = k_s [B'B'] - k_s [B''B'']. (11)$$

Based on the same adjacency arguments above, initially, $[B'B']_i = a[B]^2$ (where the *i* subscript denotes the initial time), and where a ≅ 1 based on the arguments above. Furthermore, $[B''B'']_i = 0$ since at time *s* no bond exchange has yet occurred. Moreover a mass balance gives $[B'B'] + [B''B''] = [B'B']_i = [B]^2$, where [B] is the total concentration of exchangeable bonds and is time invariant, since bond exchange events merely change the configuration of bonds in an associative network and do not alter their total number. Insertion of the mass balance into equation (13) and integration give

$$[B'B'] = \frac{1}{2}[B]^2 \left(1 + e^{-2k_s t}\right) (12)$$

Moreover, the total concentration of [B'] in this time regime is given by (since no [B'B'''] exists without diffusion) $[B'] = [B'_0] + 2[B'B']$. Furthermore, without diffusion, the concentration of $[B'_0]$ cannot change and is fixed as $[B'_0] = [B'_0]_i = [B] - 2[B'B']_i = [B] - 2[B]^2$. It follows that $[B'] = [B] - 2[B]^2 + 2[B'B']$, or $[B'] = [B] + [B]^2 (e^{-2k_s t} - 1)$. Substitution then yields, for short times at low temperatures

$$\frac{[B']}{[B]} = 1 + [B]\left(e^{-2\frac{t}{\tau_s [B]}} - 1\right) (13)$$

At long times, we model the diffusional process empirically, as a stretched exponential relaxation function as is empirically the case for glass forming liquids. This yields, for low temperatures the following time dependent bond relaxation behavior:

$$\frac{[B']}{[B]} = \left[1 + [B]\left(e^{-2\frac{t}{\tau_s [B]}} - 1\right)\right] \exp\left[-\left(\frac{t}{\tau_{diff}}\right)^\beta\right], (14)$$

where *β* is a stretching exponent for the diffusion process and $\tau_{diff}$ is the diffusional relaxation time (or subdiffusional relaxation time, see below), which is expected to scale roughly with the *α* relaxation time $\tau_\alpha$, albeit with a modestly weaker temperature dependence due to Stokes-Einstein breakdown effects that are well documented in the glass formation range.

This reaction diffusion theory thus makes several characteristic predictions for bond relaxation in associating polymers:

1) The system should exhibit a crossover from Arrhenius reaction-dominated dynamics at high temperature to super-Arrhenius diffusion-dominated dynamics at low temperature;
2) The crossover between these regimes should occur when the strand diffusion time (some multiple of the *α* relaxation time) approaches the bond exchange time;
3) In the high-temperature limit, bond relaxation should be roughly exponential in time; in the low temperature limit, bond relaxation should involve a two-step character in time, with the amplitude of the first step governed by the fraction of associable bonds that are in a locally exchangeable conformation at any given instant (a static structural property).

**Experimental Methods**

*Synthesis*

Poly(propylene glycol) bis(2-aminopropyl ether) (PPG230, PPG400, and PPG4000 with Mn=230, 400, and 4000 g/mol, respectively) was purchased from Sigma-Aldrich, and benzene-1,3,5-tricarbaldehyde (BTA, 98%) was purchased from Tokyo Chemical Industry (TCI). All samples were used as received. 2g of PPG was initially dissolved in 10 mL of Tetrahydrofuran (THF, 99.9%, anhydrous, inhibitor-free, Sigma-Aldrich). This PPG solution under stirring was added dropwise to BTA solutions in THF whose concentrations were adjusted to obtain a 1:0.75 molar ratio between the amine groups from PPG and the aldehyde groups from BTA. This mixture was reacted at 40 °C under nitrogen protection for 18-30 h. After solvent evaporating using rotary evaporation, the solution was poured into a Teflon dish and moved into a vacuum oven to further evaporate the solvent at 40 °C for 24h and then to dry at 85 °C for 48h. These systems formed crosslinked networks (Figure S3), as evident from the extended rubbery plateau (Figure S4), and have sufficient number of free $NH_2$ groups (0.5 per molecule) critical for transimination bond exchange mechanism [2].

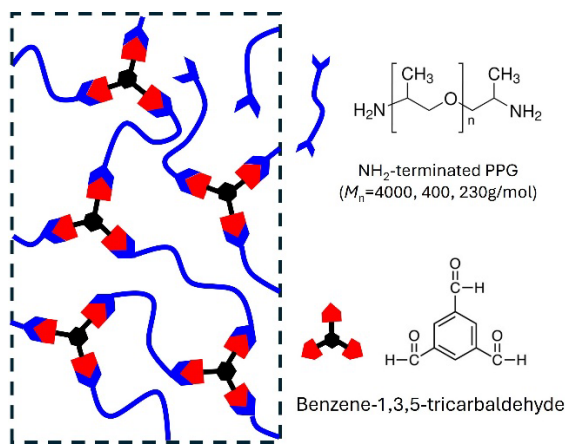

*Figure S3.* A schematic of the studied vitrimer network bearing imine bonds.

*Rheological measurements*

Small Amplitude Oscillatory Shear (SAOS) rheological measurements in the angular frequency range of $10^{-3}$-$10^2$ rad/s were conducted in linear regime with the strain-controlled mode of an AR2000ex rheometer (TA instrument) utilizing parallel plate geometry with 8mm in diameter. The sample was equilibrated before each scan to prevent temperature deviations larger than 0.2 K and all the experiments were conducted under $N_2$ flow. Before rheology measurements, the sample was dried in rheometer at 423K for 2h. The determined master curves for studied vitrimers are displayed in Figure S4. The bond relaxation time ($\tau_{bond}$) is obtained as $\tau_{bond}=1/w_{cross}$, where $w_{cross}$ is the angular frequency of G' and G''crossover.

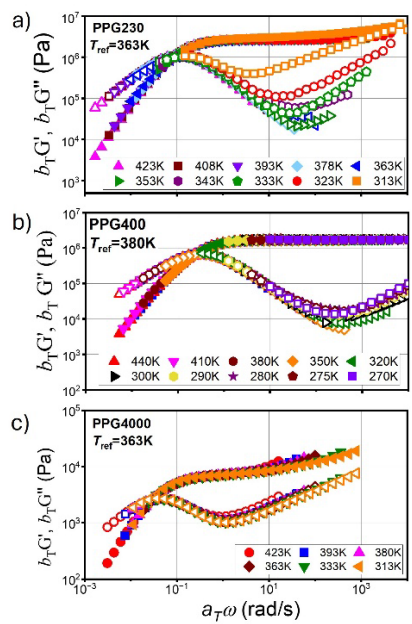

*Figure S4.* Rheological master curves of G'(solid symbols) and G'' (open symbols) for studied vitrimers.

*Broadband dielectric spectroscopy measurements*

Dielectric measurements in the frequency range from $10^{-2}$ up to $10^6$ Hz were performed using a Novocontrol GMBH Alfa impedance analyzer. In the high-frequency measurements ($10^6$ to $10^9$ Hz), we applied Agilent 4291B impedance analyzer connected with the Novocontrol GMBH system. The measurements up to $10^6$ Hz were performed in a parallel plate dielectric cell made of sapphire and invar steel with an electrode diameter of 10 mm and a capacitance of ~3.5 pF with an electrode separation of ~0.15 mm. During the measurements in a high-frequency regime, the sample was placed between two gold-plated electrodes with diameter of 6 mm and gap ~ 0.10 mm). During the measurements, the sample was equilibrated for 10 min at each temperature to reach thermal stabilization within 0.2 K. Before measurements, the sample was dried in BDS Cryosystem at 423K for 2h. The representative spectra are shown in Figure S5.

To determine the segmental/α-structural relaxation time of synthesized vitrimers, the dielectric spectra were analyzed using the Havriliak-Negami (HN) function:

$$\varepsilon^*(\omega) = \varepsilon_\infty + \sum_k \frac{\Delta\varepsilon_k}{\left(1+(i\omega\tau_{HN,k})^{\beta_k}\right)^{\gamma_k}} + \frac{\sigma_{DC}}{i\omega\varepsilon_0} + A\omega^{-b} \quad (18)$$

where $i$ is the imaginary unit, $\varepsilon^*$ is the complex permittivity, $\varepsilon_\infty$ and $\varepsilon_0$ are the dielectric constants at the infinite high frequency and the vacuum permittivity, $\omega$ is the angular frequency, $\tau_{HN,k}$, $\Delta\varepsilon_k$, $\beta_k$, and $\gamma_k$ are the characteristic HN time, dielectric amplitude, and shape parameters of the $k^{th}$ relaxation process, respectively. $\sigma_{DC}$ is the dc-conductivity, and $A$ and $b$ are constants. The characteristic relaxation time of the $k^{th}$ relaxation process can be obtained from the characteristic HN time:

$$\tau_k = \tau_{HN,k}\left[\sin\frac{\beta_k\pi}{2+2\gamma_k}\right]^{-1/\beta_k}\left[\sin\frac{\beta_k\gamma_k\pi}{2+2\gamma_k}\right]^{1/\beta_k} \quad (19)$$

Two HN functions (segmental and chain modes, bottom Figure S5) were employed to analyze the PPG4000 dielectric spectra in the studied frequency window, while one HN function was used to fit the dielectric spectra for PPG230 and PPG400 (short chains do not show well separated mode). The fitting results are shown in Figure S5 as lines.

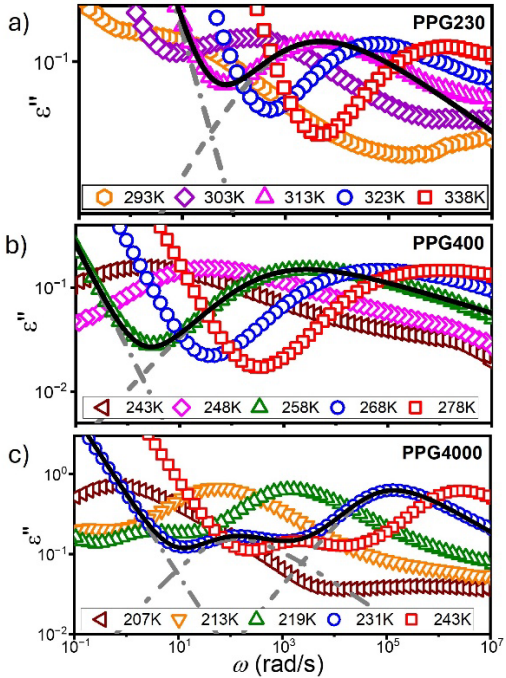

*Figure S5. Representative spectra of $\varepsilon''(\omega)$ obtained from BDS for studied vitrimers. The lines in each figure show the representative fit of the spectra to Eq. S18. The gray dash dot, dash dot dot, and dashed lines present the dc-conductivity contribution, the normal mode, and segmental relaxation mode.*

## Supplementary Experimental Data

Fits of experimental relaxation time data to equation 4 in the main text for PPG230 and PPG400 samples are shown in Figure S6. The equation 4 provides reasonable description of the data but revealed surprisingly large prefactor $c \sim 10^7 - 10^{10}$.

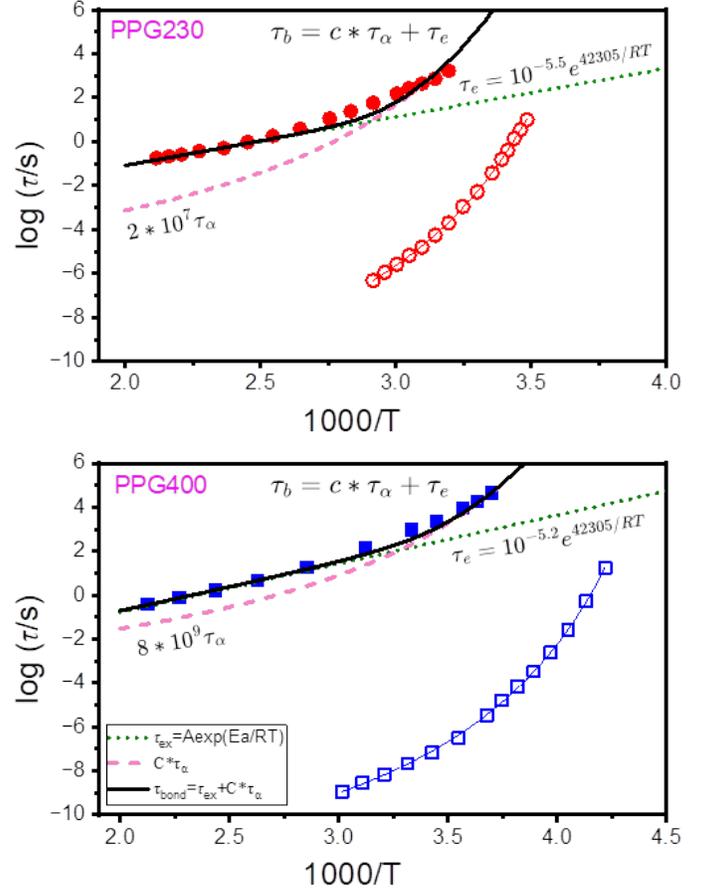

*Figure S6. Terminal (solid symbols) and segmental (open symbols) relaxation times for PPG230 and PPG400 samples. Solid lines show fits of the terminal relaxation time to the equation (4) in the main text. The dotted green line is the Arrhenius behavior with energy barrier fixed to the value estimated for $\tau_e$ for the PPG 4000 system. Thin solid lines are fits of the temperature dependence of $\tau_\alpha$ to a Vogel-Fulcher-Tamman behavior. Dashed lines present $c\tau_\alpha$ term and the numbers show the value of the prefactor c in the eq. 4.*

## Supplementary References